\documentclass{llncs}
\usepackage{llncsdoc}

\begin{document}
\title{Math-Net.Ru as a Digital Archive\\ of the Russian Mathematical Knowledge\\ from the
XIX Century to Today \thanks{The final publication is available at http://link.springer.com.}}
\titlerunning{Math-Net.Ru as a Digital Archive}
\author{Dmitry~E.~Chebukov \and Alexander~D.~Izaak \and
Olga~G.~Misyurina \and  Yuri~A.~Pupyrev \and Alexey~B.~Zhizhchenko}
\authorrunning{Chebukov, Izaak, Misyurina, Pupyrev, Zhizhchenko}
\tocauthor{D.\,E.~Chebukov, A.\,D.~Izaak, O.\,G.~Misyurina, Yu.\,A.~Pupyrev, A.\,B.~Zhizhchenko}
\institute{Steklov Mathematical Institute of the Russian Academy of Sciences}

\maketitle

\begin{abstract}
The main goal of the project Math-Net.Ru 
is to collect scientific publications in Russian and Soviet
mathematics journals since their foundation to today and the authors of
these publications into a single database and to provide access to full-text
articles for broad international mathematical community. Leading Russian
mathematics journals have been comprehensively digitized dating back to the first volumes.
\end{abstract}

\section{Introduction}
Math-Net.Ru ({\tt http://www.mathnet.ru}) is an information system developed at the
Steklov Mathematical Institute of the Russian Academy of Sciences and designed to
provide online access to Russian mathematical publications for the international
scientific community. It is a non-profit project supported by the Russian Academy of Sciences and working in the first place with journals founded by the RAS, 
but covering also other high-quality math journals. 
The project was started in 2006. 
Its main idea is to digitize the full archives of leading
Russian and Soviet mathematics journals going back to the first volumes. 
Old Russian and Soviet mathematics journals especially published before the 1930th years
could only be found in several libraries, i.e.\ in fact were hardly available to the public.

The Journals section is a key component of the system. Other sections include Persons,
Organizations, Conferences and Video Library \cite{MathNet:1}.  The mobile version of the database {\tt http://m.mathnet.ru} reproduces the most important functionality of the system but adopted for viewing on smart phones and other mobile devices.

The system has two-server architecture which includes a MSSQL database server powered by Windows 2008 server and an Apache web server powered by Linux. 
The servers are connected by a 1Gb direct network line and are located in the same server rack. All webserver scripts are written on PHP, also MSSQL stored procedures are used
in SQL logic. The database including statistics has size about 80Gb, the total size of the full-text PDF files is about 110 Gb. The total size of videofiles is 2600 Gb.

\section{Journals}
The section contains a collection of 120\,000 articles published in 86 mathematical
and physical journals. The number of journals and papers is constantly growing.
Most articles were published in Russia, but there are also journals from the former USSR: Ukraine, Belorussia, Kazakhstan, Republic of Moldova.

The page of a journal provides information about its founder, publisher and the editorial board.
The archive of the journal represents both the current issues and its historic archives, including full texts articles.
Access to full-text PDF files is specified in an agreement signed with each journal, normally access is free except for the recent (2--3 years) issues. 

We have comprehensively digitized historic archives of the leading Russian and Soviet
mathematics journals back to the fist volumes, the list includes:
\emph{Algebra i Analiz}  (since 1989); 
\emph{Zhurnal Vychislitel'no{\u i} Matematiki i Matematichesko{\u \i} Fiziki} (since 1961);
\emph{Diskretnaya Matematika} (since 1989);
\emph{Funktsional'nyi Analiz i ego Prilozheniya} (since 1967);
\emph{Bulletin de l'Acad\'emie des Sciences}  (1894--1937);
\emph{Izvestiya Akademii Nauk. Seriya Matematicheskaya} (since 1937);
\emph{Matematicheskoe Modelirovanie} (since 1989);
\emph{Matematicheskii Sbornik} (since 1866);
\emph{Matematicheskie Zametki} (since 1967);
\emph{Trudy Matematicheskogo Instituta im.\ V.\,A.~Steklova} (since 1931);
\emph{Uspekhi Matematicheskikh Nauk} (since 1936);
\emph{Teoreticheskaya i Matematicheskaya Fizika} (since 1969).

The original title, abstract and keywords, English translation title, abstract and keywords,
a link to the English version, a list of references and a list of forward links
are supplied for every paper. Titles, abstracts, keywords, references and
forward links are stored in the database in the \LaTeX{} format. We use MathJax technology ({\tt http://www.mathjax.org})
to output mathematics on the website. For every paper we provide external links
to all possible representations of the publication in Internet including links to Crossref,
MathSciNet, Zentralblatt MATH, ADS NASA, ISI Web of Knowledge, Google Scholar links to the
references cited and related papers. Enhanced search facilities include search for publications by
keywords in the title, abstract or full-text paper, by the authors' and/or institutions names.

For most journals we provide information about their citation statistics and impact factors~\cite{MathNet:2}.
Impact factors are calculated on the basis of forward links (back references) stored in the
database. English version journals are supplied with the classical Impact Factors calculated
by the Institute for Scientific Information (ISI) of the Thomson Reuters Corporation
(ISI Web of Knowledge). It is important to note that the classical (ISI) impact factors do
not include citations of the Russian versions.  
We take into account citations of both versions
and calculate the integral impact factor of the journal. This includes citations in classical
scientific journals, but also citations in conference proceedings, electronic publications.
Table~\ref{MathNet:t1} provides examples of the number of references to the volumes of years 2009--2010 and the values
of Impact Factors 2011 of some journals provided by the Math-Net.Ru and ISI Web of Knowledge. 
A significant difference between the citation numbers and
impact factors of Math-Net.Ru and ISI Web of Knowledge is explained by the fact that the
latter does not take into account references to the Russian versions of papers. Proofs of the
data stated in the Table 1 can be found in the section ``Impact factor'' of the page of the corresponding
journal on Math-Net.Ru and in Journal Citations Reports provided by ISI Web of Knowledge
system.  Our system calculates one-year and 2-year impact factors (similar to classical) and also 5-year ones.

\begin{table}[ht!]
\caption{Citation number for volumes of years 2009--2010 and Impact Factors~2011,
provided by Math-Net.Ru and ISI}
\label{MathNet:t1}
\begin{center}
\begin{tabular}{|l@{\ \vline\ }c@{\ \vline\ }c@{\ \vline\ }c@{\ \vline\ }c@{\ \vline}}
\hline
Journal &\multicolumn{2}{c}{Math-Net.Ru values}\ \vline\   &\multicolumn{2}{c}{ISI values}\ \vline
\\
\hline
& Citations &Impact&Citations &Impact
\\
&number &Factor &number &factor
\\
\hline
\emph{Matematicheskii Sbornik} 
& {\bf130} &0.813 &{\bf85} &0.567
\\
\hline
\emph{Trudy Matem. Instituta im.\ V.\,A.~Steklova}
& {\bf75} &0.455 & {\bf42} &0.171
\\
\hline
\emph{Avtomatika i Telemekhanika}
& {\bf227} &0.698 & {\bf96}	 &0.246
\\
\hline
\emph{Diskretnaya Matematika}
& {\bf43} &0.483 &-- &--
\\
\hline
\emph{Siberian Electronic Mathematical Reports}
& {\bf37} &0.378 &-- &--
\\
\hline
\emph{Russian Journal of Nonlinear Dynamics}	
& {\bf35} &0.407 &-- &--
\\
\hline
\end{tabular}
\end{center}
\end{table}

It is noteworthy that for many Russian journals ISI does not provide impact factors so
Math-Net.Ru data is a single way to evaluate the citation indexes of the journal. Table~\ref{MathNet:t1}
provides some  examples of Russian mathematical journals having no classical (ISI)
impact factor but a significant number of citations in Russian and international sources.

\section{Persons and Institutions}
Portal Math-Net.Ru also includes comprehensive information
about Russian and foreign mathematicians and  institutions
where authors of publications work or study. Up to now the database includes 52\,000
individual persons and 4\,000 institutions. Visitors of the website are free to register
online and to contribute to the database in case when they have at least one
published article in a scientific mathematics or physics journal. Personal web page provides
the list of personal publications and presentations, keywords, the list of scientific
interests and biography, web-links to additional personal resources. Special tools are available to arrange a full list of
personal publications, including papers not available within the Math-Net.Ru system.
The web-page of an institution contains general information, a link to its original web-page
and a list of authors whose papers are presented in Math-Net.Ru.

\section{Citation and Forward Link Database}
The citation database accumulates the reference lists and forward links of all
the publications available at Math-Net.Ru as well as personal lists of publications
of the authors. All references are collected into a single database and stored
in the format AMSBIB~\cite{MathNet:2} developed in the Division of Mathematics of the
Russian Academy of Sciences. The bibliography is stored using special \LaTeX{}
commands dividing a reference into several parts: journal name, authors, publication
year, volume, issue, pages information, additional information about the volume/issue
and other possible publication details. By means of these commands it is possible
to avoid manual markup of the list of references; it also enables an automatic
creation of the bibliography in PDF/HTML/XML formats, arranging hyperlinks to various
publication databases including MathSciNet, Crossref and ZentalBlatt Math.
Since all the references are collected into a single database
an advanced cited-reference search by various terms is arranged: publication
title, year, author, pages. The reference database is much wider than the database
of Math-Net.Ru publications and a search through the reference database results
in additional information about articles.

\section{Video-library, Conferences, Seminars}
Our project thoroughly collects information about mathematical events occuring
in Russia and the states of the former Soviet Union. This concerns scientific conferences
and seminars, public lectures. Most information about such events is provided
by the organizers. The system software allows arranging an event home page, which
includes general information, the list of organizers, the event schedule and
a list of presentations with links to own web pages. A presentation webpage contains
the title, abstract, date and place of the event and includes additional materials such as a list of references,
PowerPoint files and a video-record when available. We
encourage conference/seminar organizers to record videos of the presentations
and we take on post-processing of the video files. Our system accepts the most popular video formats and enables viewing them online in
all operation systems, including mobile devices. The mobile version of the system provides full access to the video-library.
The system offers viewing videos in normal and full High Definition
quality. An online access to all video records is free, all video files can be downloaded
for home viewing.

\section{Manuscript Submission and Tracking System}
The website is managed by a contents management system, which provides necessary
functionality to add/update/remove any information available. The content management
system is used to manage current and archive publications of the journals,
to communicate with authors and to create reports for editorial needs. 
The content management system resolves the problem of the creation of a document processing system in
the editorial office of a Russian scientific journal. Most western publishers provide
such systems for their journals but they cannot be used in Russian journals due to
lack of Russian language fields. The content management system includes all kinds of editorial activities from the
submission of a manuscript to the publication of the peer-reviewed paper in print and online.

The main features of the system include:
submission of a manuscript by the author in electronic form at the journal website;
registration of the authors in the database of persons;
registration of the manuscript in the paper database and arranging a paper flow process, which includes
\emph{classification},
\emph{peer review},
\emph{authors' revision},
\emph{scientific editing},
\emph{translation into English},
\emph{editing of the English version publication},
\emph{publication in print and online};
communication facilities between authors, referees, translators, typesetters, editorial board members and other people involved into publication process;
personal access of the authors, referees, editors to editorial information necessary for publication process;
arranging a list of forthcoming papers;
sending email notifications from the database;
creation of automatic reports for editorial needs.

Manuscript submission is available at the journal home page for registered authors only.
New authors should first fill in a registration form. Manuscript submission process
consists of filling in several online forms providing information about the manuscript
title, abstract, authors, keywords; then the author is asked to supply a full-text manuscript
in \LaTeX{} and PDF formats. The editor is notified about a new submission by email,
examines it on the subject of compliance with the journal rules and starts the peer-review
process. Every editorial paper flow record can be supplied with a number of documents
(files) specifying details of the process. The system generates comprehensive reports about all kinds of editorial activities.

Access to the manuscript submission system depends on the user's role in the publishing process: author, referee, editor, journal administrator.
Authors can only see paper flow details with hidden referees names. Authors are able to submit a revised version of the manuscript and download the final PDF of the published paper. 
Referees can download full texts of papers and upload reviews. They have access only to those papers they are working with. Editors normally register new papers, add paper flow records, communicate with the authors,
referees, typesetters. Journal administrators can amend anything within their journals.

\end{document}